%% file: relnr.tex
\documentstyle[aps,epsf]{revtex}
\def\bfr{\begin{flushright}}
\def\efr{\end{flushright}}
\def\i{\item}

\input macro

\begin{document}

\title{Relativistic and non-relativistic studies of nuclear matter}
\author{Manoj K. Banerjee$^{1}$ and John A. Tjon$^{1,2}$}

\address{$^1$ Department of Physics, University of Maryland, \\
College Park, MD, USA, 20742-4111\\
$^2$ Institute for Theoretical Physics,
University of Utrecht \\
3508 TA Utrecht, The Netherlands}
\date{\today}

\maketitle

\begin{abstract}

Recently we showed that while the tensor force plays an important role in
nuclear matter saturation in non-relativistic studies, it does not do so
in relativistic studies. The reason behind this is the role of $M^*$, the
sum of nucleon mass and its attractive self-energy in nuclear matter. Yet
nonrelativistic calculations at a certain level of approximation are far
less difficult than comparative relativistic calculation. Naturally the
question arises if one can modify a nonrelativistic method, say, the
lowest order Brueckner theory (LOBT), to reproduce approximately the
results of a relativistic calculation. While a many body effect, the role
of $M^*$ is intrinsically relativistic. It cannot be simulated by adding
multi-body forces in a nonrelativistic calculation. Instead, we examine if
adding a set of recipes to LOBT can be useful for the purpose.  We point
out that the differences in the results arise principally from two reasons
- first, the role of $M^*$ and second, the disappearance in a relativistic
treatment of the gap in the hole and particle energy spectra, present in
LOBT.  In this paper we show that LOBT, modified by {\it recipes} to
remove these two reasons, generates results quite close to those of
Dirac-Brueckner theory.

\end{abstract}

\section{Introduction}

In a recent paper~\cite{BT} we have shown that a relativistic treatment of the saturation
mechanism of nuclear matter differs in an essential way
from a  non-relativistic description.
As has already been pointed out by Bethe more than 40 years ago,
assuming that the nucleons behave non-relativistically,
saturation takes place 
because the large attractive contribution of second and
higher OPEP tensor force is increasingly quenched with increasing density
due to Pauli blocking. A careful analysis of the relativistic results of
Amorim and Tjon~\cite{AT} showed that the role of the tensor force in the
saturation mechanism is significantly reduced. The saturation occurs
mainly due to reduction of the scalar charge with increasing density, a
result familiar from mean field treatments of nuclear matter~\cite{SW}.

Simultaneously, it should be stressed that the role of the tensor force in
the two-nucleon problems continues to be large in a relativistic treatment
exactly as in a non-relativistic treatment. In Ref.~\cite{BT} we showed
that the reduction of the role of the tensor force is due to the role of
$M^*$, a manifestly many-body effect.

While the issue of saturation mechanism is quite dramatic there are also
detailed differences in the results of non-relativistic lowest order
Brueckner theory and its relativistic counter part - Dirac-Brueckner
theory.  This apparent shift in the physics of nuclear matter in going
from non-relativistic to relativistic description is quite surprising and
raises question about the correct treatment of the problem.  Because of
the precision and sophistication achieved in non-relativistic treatments
one cannot justifiably suggest that it be abandoned. In particular, one
may attempt to modify these equations to also include some of the salient
features of a relativistic analysis.  Results of relativistic
Dirac-Brueckner theories are closely tied to two features: a
self-consistent $M^*$ and no gap in the single particle spectrum at the
top of the Fermi sea.  We examine in this paper the possibility of
incorporating these two features in a non-relativistic theory.

\section{Modification of the Brueckner-Bethe-Goldstone equations}

Starting from the lowest order non-relativistic Brueckner-Bethe-Goldstone
(LOBT)  equations, one may attempt to incorporate as recipes the
previously mentioned features present in the relativistic Dirac-Brueckner
analysis. The existence of a gap in the single nucleon spectrum at the
Fermi surface depends on the organization of the many-body theory such as
- the non-relativistic LOBT or relativistic Dirac-Brueckner (DB) theory.
The LOBT is based on Goldstone linked cluster expansion. The propagators
between two events are global. They are not single particle propagators as
in Feynman perturbation expansion. In LOBT different sets of diagrams are
summed in the definition of the hole and the particle energies and,
naturally, a gap arises at the Fermi surface.

An alternative approach is based on the Green's function method~\cite{CM},
which is structurally the same as Feynman perturbation expansion but using
non-relativistic physics. Here the hole and the particle energies sum
similar sets of self-energy diagrams. As a natural consequence the single
particle potential is  {\it continuous across the Fermi
surface}~\cite{CM}, {\it i.e. ,} there is no gap. In DB the single nucleon 
energies are also just the self-energies and there is no gap. 

  The absence or presence of a gap can be built into the 2-nucleon
Green's function ${S}_2$ appearing in the Bethe-Goldstone equation for
the G-matrix
\beq
\langle \v p\,'\mid G \mid \v p \rangle = \langle \v
p\,' \mid\b V\mid\v p \rangle +
\sum_{\lambda,i}\int\frac{d^3 p''}{(2\pi)^3}
\ \ \langle \v p\,' \mid V\mid\v p\,''\rangle \ \
{S}_2(\v p\,'') \ \  \langle \v p\,''\mid G \mid\v p \rangle,
\l{BGE}
\eeq
where $\lambda,i$ are the helicities and isospin of the intermediate
state and ${S}_2$ is the Pauli-blocked 2-nucleon Green's function
\beq
{S}_2(\v p\,'')=\frac{Q_{Pauli}}{E(p'')-\epsilon(p)}
\eeq
with $E$ and $\epsilon$ being the single nucleon energy above and below
the Fermi momentum $p_F$
. In standard LOBT
\beq
E(p)=\frac{p^2}{2 M} ;\,\,\,
\epsilon(p)=A + \frac{p^2}{2m^*}.
\label{gap}
\eeq
The quantity $A$  and the effective mass $m^*$ are usually
determined in the LOBT calculations in a self-consistent way.

From a practical point of view, it may be useful to
explore the possibility of finding {\it recipes} which when incorporated
into non-relativistic LOBT reproduce closely the results of Dirac-
Brueckner theory. We have tried this and the next section presents the
results using our recipe of including these effects. All numerical results
reported and used in this paper are generated with a modified version of a
program written by Gruemmer~\cite{FG}.

\section{LOBT including relativized $M^*$}

 The principal features of our method of incorporating the necessary
recipes into LOBT are described below. \ben \i Introduction of $M^*$.

The precise meaning of $M^*$ in the present context is made clear in
Eq~(\ref{particle}). Specifically we choose the simple parametrization
\beq 
M^*/M=1 - \alpha\,\rho/\rho_0, 
\label{alpha} 
\eeq 
where $\rho_0=0.17
fm^{-3}$ is the density of normal nuclear matter. \i Modification of the
single nucleon energy spectrum.

We  examine  two alternatives for the modified LOBT calculation. We 
choose simple quadratic forms for the single nucleon spectrum below 
and above the Fermi sea: 

\begin{eqnarray} &&{\rm for}\,\,\,p<k_F,\, \epsilon(p)=A +
\frac{p^2}{2m^*}, \label{hole} \\ &&{\rm for}\,\,\, p>k_F,\,E(p) =
\frac{p^2}{2M^*}.  \label{particle} \end{eqnarray}

 With the spectrum as defined above the gap is 
\beq
\Delta = k_F^2 (\frac{1}{2M^*} - \frac{1}{2 m^*})- A. 
\label{Del} 
\eeq

The alternatives we examined are:  \ben \i No self-consistency, no gap.
The method of doing the nonselfconsistent is as follows. One begins by
choosing somewhat arbitrarily $m^*=M$, calculate the $G$-matrix with it
and subsequently the new hole spectrum and the new $m^*$ with it and then
stop. The absence of a gap is ensured by setting \beq A= k_F^2
(\frac{1}{2M^*} - \frac{1}{2 m^*}). \label{nogapA} \eeq \i self-consistent
$\epsilon(p)$, no gap. Here the hole spectrum is obtained
self-consistently and then A is fixed with Eq.~(\ref{nogapA}). \een \een

\begin{figure}
\epsfxsize=5in \epsfysize=6in 
\begin{center}
\hspace{0.25in} 
\epsffile{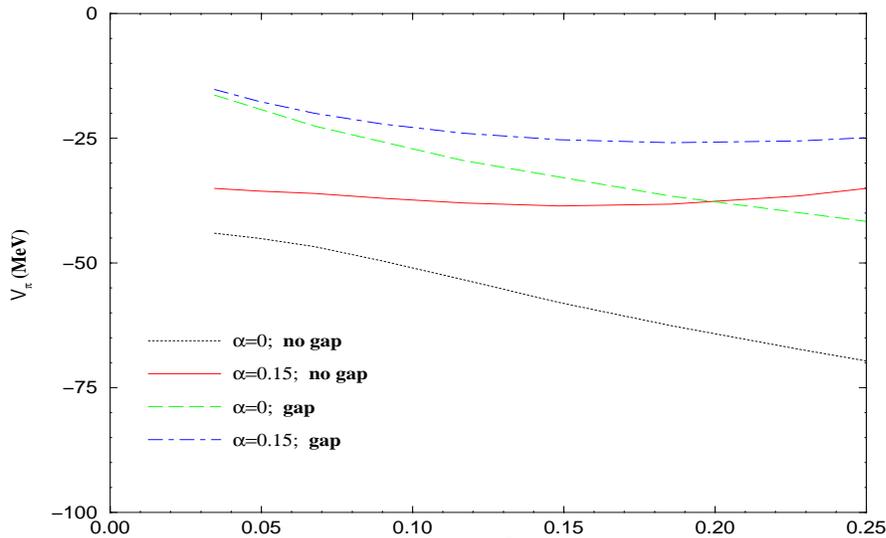} 
\end{center} 
\vspace{-2in}
\caption[F4]{ Plots of $V_\pi$ plotted as function of $\rho$ for
$\alpha=0$ and $\alpha=0.15$ and for hole energy spectrum with and without
gap.} 
\label{vopep} 
\end{figure}

A few remarks about the status of OPEP are in order.  In Ref~\cite{BT} we
have described how $M^*<M$ leads to reduction of the role of OPEP. A
measure of this role is the quantity \begin{equation}
 V_\pi = \langle NM|V_{OPEP}| NM\rangle =
g_\pi^2\frac{d}{dg_{\pi}^2}E,\label{Vpi}\end{equation} where $|NM\rangle$
is the nuclear matter ground state and $E$ is the ground state energy. In
Ref~\cite{BT} the dependence of $V_\pi$ on $M^*$ was studied by modifying the value of
the scalar self-energy $S=M^*-M$ from its self-consistently determined
value~\cite{AT} by replacing $S$ with $\alpha\,S$, $0\leq \alpha \leq 1$.
Here $M^*$ is introduced by hand and is defined by Eq.~(\ref{alpha}). It
should be noted that the definition of $\alpha$ here is different from
that used in Ref~\cite{BT}, though related in spirit.  The reduction of
$\mid V_\pi\mid$ with decreasing $M^*$ has been shown in Fig.~2 of Ref~\cite{BT}.
In Fig.~\ref{vopep} $V_\pi$ is plotted as a function of $\rho$ for
$\alpha=0$ and $\alpha=0.15$ and for hole energy spectrum with and without
gap. We see that the reduction of $\mid V_\pi\mid$ with decreasing $M^*$ occurs for
both choices of the single particle spectrum. The slight upward curvature
of $\alpha=0.15$ is due to the contribution of the first order
OPEP, being of a repulsive nature and which becomes dominant at 
higher density.

 We note that despite the reduced role in the saturation mechanism the
contribution of OPEP to $E/A$ continues to be important in the present
treatment, thus duplicating this feature of the relativistic
treatment~\cite{BT}.

\begin{figure}
\hspace{0.25in} 
\vspace{-1cm}
\epsffile{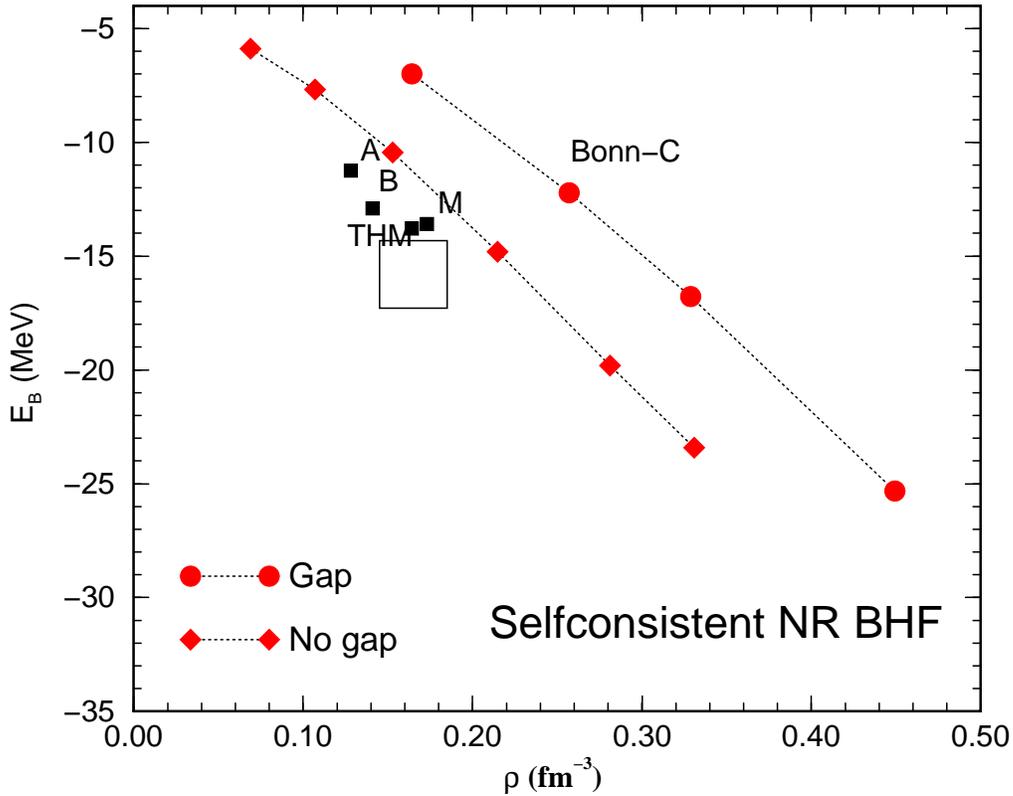} 
\vspace{0.2in} 
\caption[F4]{Predictions 
of $E_B=-E/A$ are shown for a selfconsistent calculation
(diamonds), assuming that there is no
gap present in the single particle spectrum and no $M^*$-modification.
Also is shown a selfconsistent nonrelativistic LOBT
calculation (circles). The Bonn-C potential of Ref.~\protect{\cite{M}}
is used, where the various
calculated curves are obtained by varying the $\sigma_1$- coupling
constant. For completeness, the relativistic calculations of Refs.
~\protect{\cite{AT,M,THM}}, labelled as (A, B), M  and THM, are also shown.  }
 
\label{alpha=0}
\end{figure}

In Fig.~\ref{alpha=0} we show the predictions when $\alpha=0$ and assuming
that there is no gap present in the single particle spectrum.
The curve is obtained by varying $g_{\sigma_{1}}$ parameter in the
Bonn-C potential in the range between 6.5 and 9.0. The slope
found in the case of the nonrelativistic LOBT calculation
(filled circles) agrees well with the linear Coester-line.
In accordance with~\cite{CM} we find that dropping the gap in the single
particle spectrum leads to a significant shifting towards the
empirical result. It is however not sufficient to reproduce
the empirical data. This is achieved by including the $M^*$ 
modification. It has the
important effect of changing the slope of the found linear
dependence. The predictions are shown in Fig.~\ref{nogap}.

The calculations have been carried out using the recipes
described earlier. The Bonn-C potential has been used with the
original $g_{\sigma_{1}}=8.63$ and where the $M^*$-modification is
implemented by hand using the $\alpha$-parameter of Eq.~(\ref{alpha}).
The main features to note are as follows: 
\begin{figure}
\hspace{0.25in} 
\epsffile{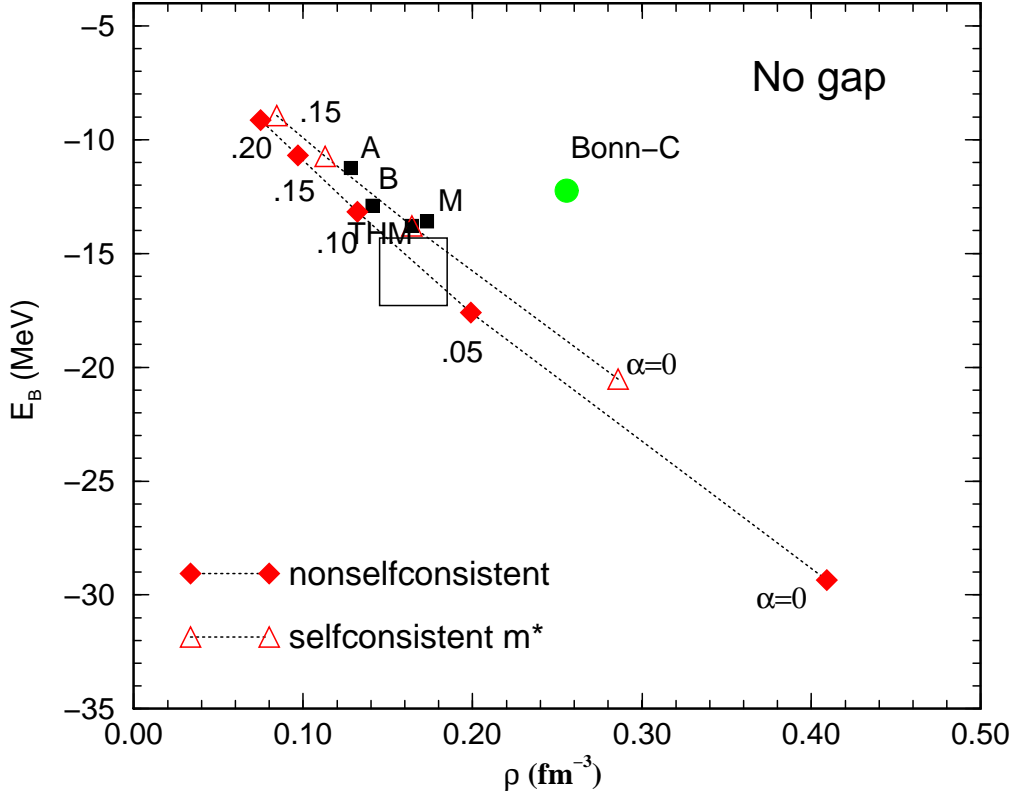} 
\vspace{0.2in} 
\caption[F4]{
Predictions of $E_B=-E/A$ assuming that there is no
gap in the single nucleon spectrum present. The two curves are obtained by varying the
$\alpha$-parameter. The results of nonselfconsistent
calculations are shown as filled diamonds. Those of selfconsistent
calculations are shown as triangles. For completeness the results of
nonrelativistic LOBT calculation for BonnC potentials of
Ref.~\protect{\cite{M}} and the relativistic calculations of Refs.
~\protect{\cite{AT,M,THM}}, labelled as (A, B), M and THM, are also shown.  }
 
\label{nogap} 
\end{figure} 

\ben
\i Once the gap is removed the results are not
all that different for non-selfconsistent and selfconsistent
calculations. 
\i Use of $\alpha>0$ is critically important to reproduce the 
empirical data. 
\i It is not surprising that the modelling which agrees
best with the relativistic results of Refs. ~\protect{\cite{AT,M,THM}}
have selfconsistent hole spectrum and no gap.  Both features are present
in the relativistic Dirac-Brueckner calculations. \i Finally, we stress
that what we have presented are recipes and not theories.  The removal of
the gap is done by hand. The quantity, $M^*$ is introduced by hand to
simulate the very important feature of relativistic many nucleon problem.  
It appears that our recipes prefer $\alpha\sim 0.15$, while Ref.~\cite{AT}
finds $\alpha \simeq 0.3$. \een

\section{Concluding remarks}

One expects a non-relativistic theory to be able to join on to a
relativistic theory in a smooth fashion.  After considerable numerical and
analytical examination of the issues we conclude, not too surprisingly,
that it is not possible to generate the two features, gap in the single
nucleon spectrum and the role of $M^*$, in a non-relativistic theory in a
natural way. Yet the level of rigor and completeness of non-relativistic
nuclear many-body theory has risen well above LOBT.  See Ref.~\cite{VRP}
for a recent review. The improvements have taken account of many-nucleon
interactions in significantly greater detail and precision. In contrast,
the relativistic nuclear many-body treatment is essentially at the level
of Dirac-Brueckner theory. So there may be considerable reluctance to give
up the non-relativistic approach in favor of a relativistic approach.
However, the $M^*$ effect, which is one of the central issues of this
paper, arises because what is energy in non-relativistic physics manifests
itself as a sum of Lorentz scalar and Lorentz vector terms in relativistic
physics. No amount of improvement in the dynamical aspects of a
non-relativistic theory can address this issue. In other words, there is
no magic three- or more-body force which, upon inclusion in
non-relativistic calculations will imitate the role of $M^*$. In any case,
if a genuinely new force is introduced in a non-relativistic treatment a
similar force must also be be added in the relativistic treatment, unless
it is already present. Thus a truly satisfactory resolution of the problem
appears to be impossible.

Despite these caveats it is useful, from a practical point of view, to
explore the possibility of finding {\it recipes} which when incorporated
into non-relativistic LOBT reproduce closely the results of Dirac-
Brueckner theory. This we have done in the present paper
From this study we have found that the presence of a gap, as
predicted by the standard LOBT analysis, or absence has an
important effect. In addition, the $M^*$ modification changes the
slope of the nonrelativistic Coester-line significantly, leading to 
results in accordance with the Dirac-Brueckner analysis. However it 
should be stressed, that the prescriptions we have used
are simply recipes to simulate some of the aspects of the 
relativistic analysis.

\end{document}

%% file: macro.tex
\def\beq{\begin{equation}}
\def\eeq{\end{equation}}
\def\ber{\begin{eqnarray}}
\def\eer{\end{eqnarray}}
\def\l {\label}

\def\ben{\begin{enumerate}}
\def\een{\end{enumerate}}
\def\bei{\begin{itemize}}
\def\eei{\end{itemize}}
  
\def\i{\item}
\def\beq{\begin{equation}}
\def\eeq{\end{equation}}
\def\l{\label}
\def\ber{\begin{eqnarray}}
\def\eer{\end{eqnarray}}
\def\v#1{\vec{#1}}